\documentclass[twocolumn,showpacs,preprintnumbers,amsmath,amssymb]{revtex4}
\usepackage{tabularx,graphicx}

\usepackage{color}
\usepackage{hyperref}
\hypersetup{
    colorlinks=true,
    linkcolor=blue,
    filecolor=magenta,      
    urlcolor=blue,
}

\begin{document}
%\documentstyle[aps]{revtex}
%\documentstyle[preprint,aps]{revtex}
%\begin{document}

\newcommand{\beq}{\begin{equation}}
\newcommand{\eeq}{\end{equation}}
\newcommand{\beqn}{\begin{eqnarray}}
\newcommand{\eeqn}{\end{eqnarray}}
\newcommand{\bmath}{\begin{subequations}}
\newcommand{\emath}{\end{subequations}}
\newcommand{\bra}[1]{\langle #1|}
\newcommand{\ket}[1]{|#1\rangle}

%\draft
\title{Dynamics of the normal-superconductor phase transition and the puzzle of the  Meissner effect}
\author{J. E. Hirsch }
\address{Department of Physics, University of California, San Diego,
La Jolla, CA 92093-0319}

\begin{abstract} 
The analisis of Pippard \cite{pip} for the growth of the normal phase into the superconducting phase in the presence of a 
magnetic field $H>H_c$ is applied in reverse to  the case $H<H_c$ ($H_c=$critical magnetic field). We carry out the analysis both for a planar and a cylindrical geometry. As
the superconducting phase grows into the normal phase, a supercurrent is generated at the superconductor-normal phase
boundary that flows in direction opposite to the Faraday electric field resulting from the moving phase boundary.
This supercurrent motion is in direction opposite to what is dictated by the Lorentz force on the current carriers,
and in addition requires that mechanical momentum of opposite sign be tranferred to the system as a whole to ensure momentum
conservation. In the cylindrical geometry case, a macroscopic torque of unknown origin acts on the body as a whole as the magnetic field
is expelled.
We argue that the conventional BCS-London theory of superconductivity
 cannot explain these facts, and that
as a consequence the Meissner effect remains unexplained within the conventional theory of superconductivity.
We propose that the Meissner effect can only be understood by assuming that there is 
{motion of charge in direction perpendicular to the normal-superconductor phase boundary} and point out that the unconventional
theory of hole superconductivity describes this physics.
\end{abstract}
\maketitle

   \begin{figure}
 \resizebox{8.5cm}{!}{\includegraphics[width=6cm]{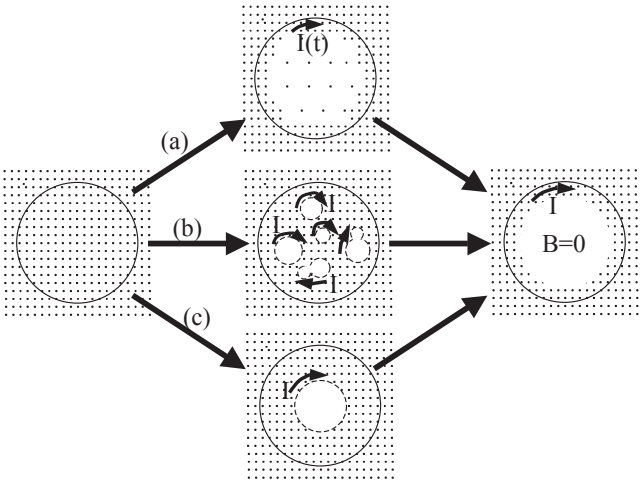}}
 \caption { Three possible routes for the magnetic field expulsion in a cylindrical superconductor. 
 The dots represent magnetic field lines coming out of the paper. The arrows give the 
 direction of the currents ($I$). See text for a discussion of the processes. }
 \label{figure1}
 \end{figure} 
\section{introduction}
When a metal is cooled into the superconducting state in the presence of a magnetic field H, the magnetic field is expelled from the interior.  A current flows within a London penetration depth ($\lambda_L$) of
the surface that generates a magnetic field that exactly compensates the magnetic field in the interior. This is the Meissner effect, discovered experimentally in 1933 \cite{meissner}. In this paper we
analyze the dynamics of this process and argue that it cannot be understood within the conventional BCS-London theory of superconductivity \cite{londonbook,tinkham}, contrary to what is generally believed.

First let us discuss some general features of the process. Figure 1 shows three conceivable routes to get from the initial to the final state of a long cylinder cooled into the superconducting state
in the presence of a uniform magnetic field along the cylinder axis. For simplicity we assume translational invariance in the direction of the cylinder axis (long cylinder). 
In route (a), a current   develops near the surface of the cylinder that gradually and uniformly depresses the magnetic field in the interior, until the magnetic field vanishes. In terms of magnetic field lines,
their density decreases uniformly in the interior of the cylinder. In (b), small superconducting domains nucleate at random positions in the sample (determined by imperfections or
temperature fluctuations), each with its own surface current that cancels the magnetic field in its interior. These domains grow and coalesce with each other, when domains
coalesce their interior currents cancel out and only the boundary currents remain. Eventually one single domain that occupies the entire sample results. In route (c), a single domain
grows from the center, with a surface current that nullifies the magnetic field in its interior, and its  boundary expands radially outward until it reaches the boundary of the cylinder.

Note that in this cylindrical geometry the magnitude of the currents around the domains in the scenarios (b) and (c) is always the same, assuming it flows within a London penetration depth
of the surface of the domain, since it always nullifies the magnetic field in the interior.

We argue that the scenario (a) is untenable on physical grounds. In this scenario, while the transition is taking place a non-zero magnetic field exists throughout the interior preventing the
establishment of phase coherence in the interior region, hence preventing formation of the superconducting phase that would lower the system's free energy. In the absence of
condensation energy there is no source of energy to drive the surface current and pay the electromagnetic energy price involved in changing the magnetic field in the interior.  
Thus  scenario (a) can be discarded on purely theoretical grounds. 

Experimentally, there is evidence that the transition can occur through the process described in (b) \cite{nucl} and also in (c), if a slightly smaller magnetic field is imposed at the center of the sample \cite{faber2}.
We argue that there is no fundamental difference in the processes (b) and (c) and that to understand the conceptual issues of interest here it is sufficient to concentrate on the scenario (c) 
which is what we will do for the rest of this paper.

In a seminal paper in 1950, Pippard \cite{pip} analyzed the process of growth of the normal into the superconducting phase when a magnetic field larger than a critical field is applied, as shown
schematically in Fig. 2(a). He argued persuasively that the rate of propagation of the phase boundary is governed by electromagnetic processes, and calculated the speed at which the
boundary propagates, in approximate agreement with experiment. The applied field is $H>H_c$ and it is reduced to $H_c$ at the normal-superconductor boundary through the current generated
due to Faraday's law. Here we will apply the same reasoning to the reverse process shown in Fig. 2(b), i.e. the Meissner effect. 
We will find that understanding the dynamics of this process requires making assumptions that are incompatible with the conventional BCS-London theory of superconductivity.
For completeness, we repeat Pippard's analysis of the process (a) in the next section, 
before proceeding to the case of interest (b).

 \begin{figure}
 \resizebox{8.5cm}{!}{\includegraphics[width=6cm]{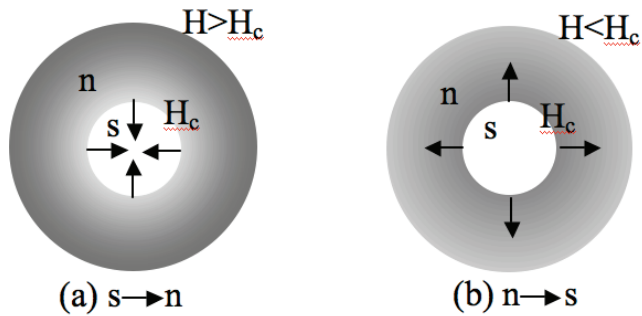}}
 \caption {(a) Growth of the normal  (n) phase into the superconducting (s) phase for a cylinder, under application of a magnetic field $H$ larger than the critical field $H_c$. 
 (b) The reverse process (Meissner effect), under an applied magnetic field $H$ smaller than the critical field. For both cases, the magnetic field at the 
 normal-superconductor boundary is the critical field $H_c$. The shading of grey indicates the magnitude of magnetic field, white color indicates no magnetic field. }
 \label{figure1}
 \end{figure}

There have been more recent theoretical studies of the growth of the superconducting phase into the normal phase \cite{dorsey, goldenfeld}, 
taking into account the possibility that the planar interface may become unstable and more complicated growth patterns develop, and 
studying the differences between type I and type II superconductors.
However these treatments do not address the fundamental questions of interest here. We will return to this point in  a
later section of this paper. The treatment in this paper applies to type I superconductors.

 \begin{figure}
 \resizebox{8.5cm}{!}{\includegraphics[width=6cm]{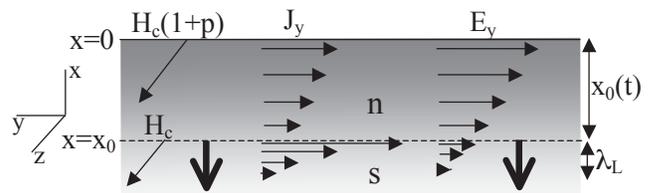}}
 \caption {Growth of the normal  (n) into the   superconducting (s) phase for a planar geometry, under application of a magnetic field $H=H_c(1+p)$ , with $p>0$.
The magnetic field points  out of the paper, the shading of grey indicates schematically its intensity. The n-s phase boundary (dashed line) moves down. }
 \label{figure1}
 \end{figure}

\section{growth of the normal into the superconducting phase}

Instead of a cylindrical geometry we will consider  the planar geometry shown in Fig. 3, following Pippard \cite{pip}. The physics is the
same and the calculation is considerably simpler. Later  in sect. IX we return to the more interesting case of a cylindrical geometry.

The applied magnetic field 
(in the $\hat{z}$ direction)  at the boundary of the sample ($x=0$)  is given by
\bmath
\beq
H(x=0)=H_c(1+p)
\eeq
with $p>0$. As the phase boundary moves in the negative $x$ direction, eddy currents are generated that generate a magnetic field
opposite to the applied one that reduces its magnitude, thus limiting the speed at which the boundary
moves into the superconducting phase. The  phase boundary is 
at the point $x=x_0(t)$ where the magnetic field is exactly the critical field $H_c$. 
\beq
H(x_0)=H_c
\eeq
\emath
In the normal phase ($x_0<x<0$), the equations to be considered are
\bmath
\beq
\vec{\nabla}\times \vec{H}=\frac{4\pi}{c}\vec{J}==>
\frac{\partial H}{\partial x}=-\frac{4\pi}{c}J_y
\eeq
\beq
\vec{\nabla}\times \vec{E}=-\frac{1}{c}\frac{\partial\vec{H}}{\partial t}==>\frac{\partial E_y}{\partial x}=-\frac{1}{c}
\frac{\partial H}{\partial t}
\eeq
\beq
\vec{J}=\sigma \vec{E}==> J_y=\sigma E_y
\eeq
\emath
Here, $\vec{H}=H\hat{z}$, $\vec{J}=J_y\hat{y}$, $\vec{E}=E_y \hat{y}$. 
Application of Faraday's law
using a contour with one edge at the phase boundary and the other deep into the superconducting phase
where $H(x)=0$ yields
\bmath
\beq
E_y(x=x_0)=\frac{H_c}{c}\frac{\partial x_0}{\partial t}.
\eeq
and replacing $E_y$ by Eq. (2c) and using Eq. (2a) the condition Eq. (3a) is
\beq
\frac{\partial H}{\partial x} )_{x=x_0}=-\frac{4\pi \sigma}{c^2}H_c \frac{\partial x_0}{\partial t}.
\eeq
\emath

Taking the x-derivative of Eq. (2a) and replacing the right-hand side by Eqs. (2b), (2c) leads to
\beq
\frac{\partial^2H}{\partial x^2}=\frac{4\pi \sigma}{c^2}\frac{\partial H}{\partial t} .
\eeq
We define
\bmath
\beq
y=\frac{x}{x_0}
\eeq
\beq
H=H_c(1+f(y))
\eeq
\emath
and from Eq. (4), $f$ satisfies the equation
\beq
\frac{\partial^2f}{\partial y^2}=\frac{4\pi \sigma}{c^2}x_0^2\frac{\partial f}{\partial t} .
\eeq
Expressing the time derivative of $f$ in Eq. (4) in terms of the time derivative of $x_0$ yields the equation
\beq
\frac{f''(y)}{f'(y)}=-\frac{2\pi\sigma}{c^2}y  \frac{\partial (x_0(t))^2}{\partial t}.
\eeq
from which we conclude that the right-hand side is independent of time, hence $x_0^2$ increases linearly
with time. We write
\beq
x_0(t)^2=\frac{\alpha p c^2}{2\pi \sigma}t
\eeq
and the parameter $\alpha$ is determined by the boundary conditions, which are
\bmath
\beq
f(y=0)=p
\eeq
\beq
f(y=1)=0
\eeq
\beq
\frac{\partial f}{\partial y} )_{y=1}=-\frac{2\pi \sigma}{c^2} \frac{\partial (x_0(t))^2}{\partial t}.
\eeq
\emath
Eq. (9c) follows from Eq. (3b).

Replacing Eq. (8) into Eq. (7) yields
\beq
f''+\alpha p y f'=0
\eeq
hence
\bmath
\beq
f'(y)= - \alpha p e^{\frac{\alpha p}{2}(1-y^2})
\eeq
\beq
f(y)=p-\alpha p \int_0^ydy'e^{\frac{\alpha p}{2}(1-y'^2})
\eeq
\emath
and $\alpha$ is determined by the condition
\beq
 \alpha \int_0^1dy'e^{\frac{\alpha p}{2}(1-y'^2})=1 .
\eeq
For small $p$, $\alpha \sim 1$ and it decreases as $p$ increases, as shown in Fig. 4. Note that Eq. (8) implies that
the higher the conductivity of the normal metal the slower the motion of the phase boundary,
as one would expect.

 \begin{figure}
 \resizebox{8.5cm}{!}{\includegraphics[width=6cm]{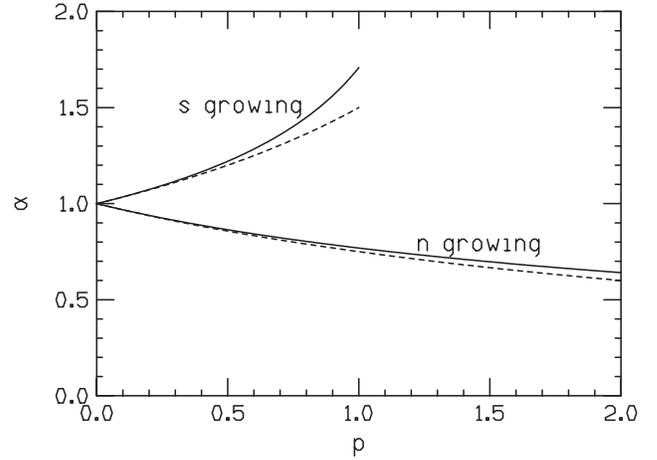}}
 \caption {Parameter $\alpha$ determined by the condition Eq. (12) for growth of the normal into the superconducting phase, and for  growth of the 
 superconducting into the normal phase determined by the condition Eq. (29) of Sect. III. The dashed lines are the values resulting from the linear approximation discussed in Sect. IX.}
  \label{figure1}
 \end{figure}

The current  and electric field in the normal phase are hence given by
\bmath
\beq
J_y=\frac{c}{4\pi}    \frac{\alpha p}{x_0} H_ce^{\frac{\alpha p}{2}(1-y^2)}=
\frac{\sigma}{c}\frac{\partial x_0}{\partial t}H_ce^{\frac{\alpha p}{2}(1-y^2)}
\eeq
\beq
E_y=\frac{c}{4\pi\sigma}    \frac{\alpha p}{x_0} H_ce^{\frac{\alpha p}{2}(1-y^2)}=
\frac{H_c}{c}\frac{\partial x_0}{\partial t}e^{\frac{\alpha p}{2}(1-y^2)}
\eeq
\emath
and the magnetic field is given by Eqs. (5b) and (11b). $H$, $J_y$ and $E_y$ all decrease in going
from the surface of the material  ($y=0$) to the phase boundary ($y=1$).

In the superconducting phase Eq. (2a) is of course satisfied which we reproduce here for convenience,
\bmath
\beq
\vec{\nabla}\times \vec{H}=\frac{4\pi}{c}\vec{J}==>
\frac{\partial H}{\partial x}=-\frac{4\pi}{c}J_y
\eeq
and the London equation \cite{londonbook} is satisfied
\beq
\vec{\nabla}\times \vec{J}=-\frac{c}{4\pi\lambda_L^2}\vec{H}==>
\frac{\partial J_y}{\partial x}=-\frac{c}{4\pi \lambda_L^2}H
\eeq
\emath 
which upon combining  leads to
\bmath
\beq
\frac{\partial ^2 H}{\partial x^2}=\frac{1}{\lambda_L^2} H
\eeq
\beq
\frac{\partial ^2 J_y}{\partial x^2}=\frac{1}{\lambda_L^2} J_y
\eeq
\emath
hence
\bmath
\beq
H(x)=H_ce^{(x-x_0)/\lambda_L}
\eeq
\beq
J_y(x)=J_y(x_0^-)e^{(x-x_0)/\lambda_L}
\eeq
\emath
The current at the phase boundary on the superconducting side is, from Eq. (14a)
\beq
 J_y(x_0^-)= -\frac{c}{4\pi \lambda_L}H_c 
\eeq
while on the normal side of the phase boundary it is given by, from Eq. (13a)
\beq
J_y(x_0^+)= \frac{c}{4\pi}    \frac{\alpha p}{x_0} H_c
\eeq
so that it is discontinuous. The electric field in the superconducting side is obtained from Faraday's law Eq. (2b),
the expression for the magnetic field Eq. (16a) and the time derivative of $x_0$ Eq. (8)
\beq
E_y(x)=\frac{H_c}{c}\frac{\partial x_0}{\partial t} e^{(x-x_0)/\lambda_L}
\eeq
Note that the electric field is continuous at the phase boundary (cf Eq. (13b)), as expected.

Upon taking the time derivative of Eq. (16b) and using Eqs.  (17) and (19) we find
\beq
\frac{\partial J_y}{\partial t}=\frac{c^2}{4\pi \lambda_L^2}E_y
\eeq
which has a simple interpretation. With $J_y=n_sqv_y$, with $n_s$ the density of superconducting carriers
of charge $q$, $v_y$ their velocity, and using the standard expression for the London penetration depth \cite{tinkham}
\beq
\frac{1}{\lambda_L^2}=\frac{4\pi n_s q^2}{m c^2}
\eeq
with $m$ the mass of the carrier,
Eq. (20) is
\beq
\frac{\partial v_y}{\partial t}=\frac{q}{m}E_y
\eeq
which describes free acceleration of carriers of charge $q$ and mass $m$ due to the electric field
generated by Faraday's law due to the moving phase boundary.

The same analysis can be applied to a cylindrical geometry, except that in that case an analytic
solution is no longer possible. However an approximate treatment which is accurate for not
too large values of $p$ shows that the results are very similar to the results discussed in 
this section \cite{pip}.

Thus, the physics of this process, shown qualitatively in Fig. 3 and described quantitatively by Eqs. 
(8)-(22), is clear and intuitive. The moving phase boundary changes the magnetic flux and generates
the electric field $E_y$ both in the normal and the superconducting regions. In the normal region the
electric field produces eddy currents, and in the superconducting region it produces a freely
accelerating current. At the phase boundary the current drops in magnitude by a large amount
in going from the superconducting to the normal side.
This is of course because scattering processes set in, described by the resistivity $\rho=1/\sigma$.
When the boundary moves slightly down, the supercurrent at the boundary 
drops discontinuously as scattering processes
set in, and its  momentum is transmitted to the solid as a whole. 

Instead, as we discuss in the next sections, there is no similarly simple physical interpretation for the reverse process,
when the superconducting phase grows into the normal phase.

\section{growth of the superconducting into the normal phase}
We now apply the same analysis to the reverse process, where the applied magnetic field is lower
than the critical field and hence the superconducting phase grows into the normal phase,
as shown in Fig. 5. Again the 
kinetics is determined by electromagnetic processes. As the phase boundary moves into the normal 
phase, the current enhances the magnetic field at the phase boundary, which is again the critical
field $H_c$. As before, the speed of motion of the phase boundary is determined by the eddy currents
generated in the normal phase.

   \begin{figure}
 \resizebox{8.5cm}{!}{\includegraphics[width=6cm]{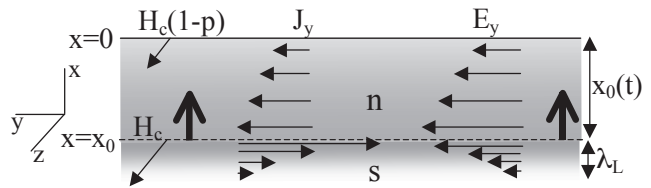}}
 \caption { Growth of the superconducting  (s) into the   normal (n) phase for a planar geometry, under application of a magnetic field $H=H_c(1-p)$ with $0<p<1$.
The magnetic field points out of the paper, the shading of grey indicates schematically its intensity. The n-s phase boundary (dashed line) moves up. }
 \label{figure1}
 \end{figure}

The boundary conditions are now
\bmath
\beq
H(x=0)=H_c(1-p)
\eeq
\beq
H(x=x_0)=H_c
\eeq
\emath
with $p>0$. Eqs. (3), (4) and (5a) still hold, Eq. (5b) gets replaced by
\beq
H=H_c(1-f(y)) .
\eeq
Eqs. (6) and (7) still hold, and Eq. (8) is replaced by
\beq
x_0(t)^2= R^2 -\frac{\alpha p c^2}{2\pi \sigma}t .
\eeq
assuming the initial distance between the phase boundary and the surface of the
material  is $R$.
Eqs. (9a) and (9b) still hold, and Eq. (9c) gets replaced by
\beq
\frac{\partial f}{\partial y} )_{y=1}=\frac{2\pi \sigma}{c^2} \frac{\partial (x_0(t))^2}{\partial t}.
\eeq
Eqs. (10), (11) and (12) get replaced by
  \beq
f''-\alpha p y f'=0
\eeq
\bmath
\beq
f'(y)= - \alpha p e^{\frac{\alpha p}{2}(y^2-1})
\eeq
\beq
f(y)=p-\alpha p \int_0^ydy'e^{\frac{\alpha p}{2}(y'^2-1})
\eeq
\emath
and $\alpha$ is determined by the condition
\beq
 \alpha \int_0^1dy'e^{\frac{\alpha p}{2}(y'^2-1})=1 .
\eeq
For small $p$, $\alpha \sim 1$, now as $p$ increases $\alpha$ also increases, as shown in Fig. 4.

The current  and electric field in the normal phase are  given by
\bmath
\beq
J_y=-\frac{c}{4\pi}    \frac{\alpha p}{x_0} H_ce^{\frac{\alpha p}{2}(y^2-1)}=
\frac{\sigma}{c}\frac{\partial x_0}{\partial t}H_ce^{\frac{\alpha p}{2}(y^2-1)}
\eeq
\beq
E_y=-\frac{c}{4\pi\sigma}    \frac{\alpha p}{x_0} H_ce^{\frac{\alpha p}{2}(y^2-1)}=
\frac{H_c}{c}\frac{\partial x_0}{\partial t}e^{\frac{\alpha p}{2}(y^2-1)}
\eeq
\emath
and the magnetic field is given by Eqs. (24) and (28b). $H$, $J_y$ and $E_y$ all increase in going
from the surface of the material  ($y=0$) to the phase boundary ($y=1$).

In the superconducting phase Eqs. (14) and (17) still hold. In particular, the current on the
superconducting side of the phase boundary is given by Eq. (17) which we reproduce here for
convenience
\beq
 J_y(x_0^-)= -\frac{c}{4\pi \lambda_L}H_c 
\eeq
while on the normal side it is given by
\beq
J_y(x_0^+)= -\frac{c}{4\pi}    \frac{\alpha p}{x_0} H_c
\eeq
so the currents at the boundary now run in opposite directions, as shown schematically in Fig. 5. 
The electric field on the superconducting side is still given by Eq. (19),
\beq
E_y(x)=\frac{H_c}{c}\frac{\partial x_0}{\partial t} e^{(x-x_0)/\lambda_L}
\eeq
which now points in opposite direction to Eq. (19) because the boundary is moving in opposite
direction. The electric and magnetic fields are of course again continuous at the phase boundary.
In contrast to the previous case where the electric and magnetic fields decrease continuously
in going from the normal to the superconducting phase, here both the electric and magnetic
fields achieve their maximum values at the phase boundary and decrease from there
both in going into the normal and into the superconducting phase.

Note that Eqs. (20) and (22) still hold here, i.e. 
\bmath
\beq
\frac{\partial J_y}{\partial t}=\frac{c^2}{4\pi \lambda_L^2}E_y
\eeq
\beq
\frac{\partial v_y}{\partial t}=\frac{q}{m}E_y
\eeq
\emath
however they have a different physical interpretation that in the previous case.
Here, Eq. (34) describes the {\it slowing down} of the carriers in the superconducting
phase  by the Faraday field that applies a force in direction opposite to the motion of
the carriers. The origin of the motion of the charge carriers and resulting current in the
superconducting phase is $not$ described by the electromagnetic processes discussed
in this section, in contrast to the situation in the case of the growing normal phase.
We discuss the situation in detail in the next sections.

  \begin{figure}
 \resizebox{8.5cm}{!}{\includegraphics[width=6cm]{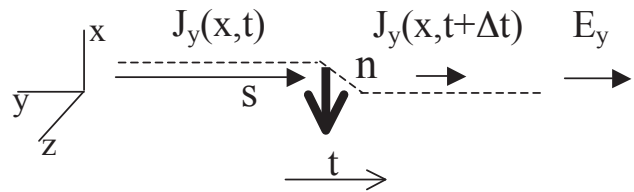}}
 \caption { Transition from the superconducting state to the normal state in a boundary layer. The charge carriers change their speed from a large
 $v_s$ to a small $v_n$ given by Eqs. (35). }
 \label{figure1}
 \end{figure}

\section{energy and momentum conservation in the s $\rightarrow$ n transition}

Figure 6 shows schematically a boundary layer that changes from superconducting to normal in a time interval $\Delta t$. The carriers in the layer of thickness
$\delta=\dot{x}_0(t)dt$ change their speed from a large $v_{s}$ when they are in the superconducting phase to a small $v_{n}$ when they become normal carriers, namely
(from Eqs. (17) and (18))
\bmath
\beq
v_{s}=-\frac{c}{4\pi \lambda_L} \frac{H_c}{qn_s}
\eeq
\beq
v_{n}=\alpha p \frac{c}{4\pi x_0} \frac{H_c}{qn_s} 
\eeq
\emath
hence there is a change of kinetic energy and of momentum per carrier, given by
\bmath
\beq
\Delta \epsilon_k=\frac{1}{2}mv_{n}^2-\frac{1}{2}mv_{s}^2
\eeq
\beq
\Delta p_y=m(v_{n}-v_{s}) .
\eeq
\emath
How are these changes compensated?

     \begin{figure}
 \resizebox{5.5cm}{!}{\includegraphics[width=6cm]{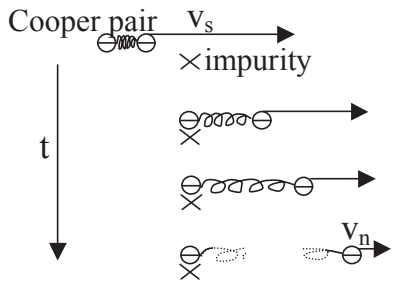}}
 \caption { Schematics of the transition shown in Fig. 6.  A Cooper pair is moving with a high center of mass speed
  $v_s$ initially. One of its
 members gets attached to an impurity, and as the pair breaks up the original kinetic energy of the pair is used up to
 pay for the binding energy. When the pair unbinds essentially all the initial  kinetic energy was used up in 
 the energy required to unbind the
 pair. In the process, the momentum of the pair is transferred to the impurity and hence to the lattice as a whole. }
 \label{figure1}
 \end{figure}

The kinetic energy of the supercurrent is $not$ dissipated in collisions, since the transition is reversible \cite{londonh}. The lowering of kinetic energy occurs because the carriers go from the superconducting to the normal phase, hence have to pay the price of the superconducting condensation energy.
The change in kinetic energy per unit volume is, neglecting the very small kinetic energy in the normal phase
\beq
\Delta E_k=n_s \Delta \epsilon_k= -n_s \frac{1}{2}mv_{s}^2=-\frac{H_c^2}{8\pi}
\eeq
where we have used Eq. (35a) for $v_s$ and  Eq. (21) for $\lambda_L$.  This is precisely the condensation energy per unit volume.  We can think of the condensation energy as the binding energy of
Cooper pairs. To unbind the pairs requires to pay this energy cost, and it is paid by the kinetic energy of the carriers which consequently slow down to
essentially zero kinetic energy. The process is shown schematically in Fig. 7. It involves interaction with
some impurity or defect, and in the process the momentum originally carried by the pair is transferred
to the lattice as a whole, with no irreversible loss of energy. 

Note also that the origin of the carrier's velocities $v_s$ and $v_n$ in Eq. (35)  is well understood. $v_s$ originates in the action of the electric field $E_y$ on the superconducting carriers. For $x$ deep into
the superconducting region the carriers are initially at rest. As the phase boundary approaches to a distance of order $\lambda_L$ the electric field starts to grow and drives the carriers
according to Eq. (22). Assuming at time $t=0$ the phase boundary is at $x_0=0$ and at time $t_0$ it  reaches the position $x$, i.e. $x_0(t_0)=x$, we have for the carriers' speed at $x$
\beq
v_y (x,t_0)=\int_0^{t_0} \frac{q}{m}E_y(x,t)=-\frac{qH_c \lambda_L}{mc}(1-e^{x/\lambda_L})
\eeq
where we used Eq. (19) for the electric field.  The term $e^{x/\lambda_L}$ is negligible for $|x|/ \lambda_L>>1$ and Eq. (38) is the critical velocity Eq. (35a). 
On the normal side, the velocity at the boundary is simply determined by the Faraday electric field and Ohm's law.

As we discuss in the next section, there is no comparable understanding of the reverse process
whereby a normal layer becomes superconducting.

    \begin{figure}
 \resizebox{8.5cm}{!}{\includegraphics[width=6cm]{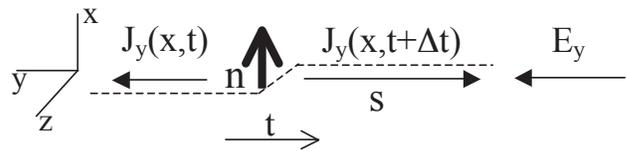}}
 \caption { Transition from the normal  state to the superconducting state in a boundary layer. The charge carriers change their speed from a small 
 $v_n$ in the positive $y$ direction (assuming $q>0$) to a large  $v_s$ in the negative $y$ direction, given by Eqs. (39). }
 \label{figure1}
 \end{figure}

\section{energy and momentum conservation in the n $\rightarrow$ s transition}

In the reverse process, where a boundary layer changes from normal to superconducting,
the direction of the current changes when the layer becomes superconducting, as calculated in Sect. III and shown in Fig. 8. Hence the direction of the
motion of the carriers reverses, and in addition the speed of the carriers increases by a large amount, 
from $v_n$ to $v_s$, given by
\bmath
\beq
v_n= -\alpha p\frac{c}{4\pi x_0}    \frac{H_c}{qn_s}
\eeq
\beq
v_s= -\frac{c}{4\pi \lambda_L}\frac{H_c}{qn_s} .
\eeq
\emath
The change of kinetic energy and of momentum per carrier are given by
\bmath
\beq
\Delta \epsilon_k=\frac{1}{2}mv_{s}^2-\frac{1}{2}mv_{n}^2
\eeq
\beq
\Delta p_y=m(v_{s}-v_{n}) .
\eeq
\emath
For the change in kinetic energy per unit volume we have
\beq
\Delta E_k=n_s \Delta \epsilon_k= n_s \frac{1}{2}mv_{s}^2=\frac{H_c^2}{8\pi}
\eeq
just like in the previous case ($s\rightarrow n$ transition) but with opposite sign. The increase in kinetic energy   is supplied by the condensation energy.
Again the process is reversible \cite{londonh}.

There are several questions that need to be answered:

(1) What is the physical origin of the speed $v_s$, Eq. (39b)?

(2) How is the momentum change given by Eq. (40b) compensated?

(3) What is the physical mechanism  by which the condensation energy becomes kinetic energy of the supercurrent, as given by Eq. (41)?

We argue that the conventional BCS-London theory of superconductivity offers no answers to these questions. Let us discuss each of these questions.

\subsection*{(1) What is the physical origin of $v_s$?}

The Faraday field $E_y$ (Eq. (30b) or (33)) points in the positive $y$ direction (see   Fig. 8). Carriers in the normal state move in the positive $y$ direction, driven by $E_y$ (assuming $q>0$), and when the boundary layer becomes superconducting
the carriers suddenly  reverse their direction and move with large speed $v_s$ in the negative $y$ direction, opposing the force $qE_y$ that drives  them to move in the
positive $y$  direction.

How does the condensation process, or the Cooper pair formation process, make the carriers acquire a center of mass momentum in direction opposite to the applied electric force?
Furthermore, this center of mass momentum is only acquired if a magnetic field is present: if a material goes superconducting in the absence of a magnetic field, no $v_s$ is
acquired by the carriers condensing into the superconducting state.

Perhaps it will  be argued that in the process of becoming part of the superconducting condensate and establishing phase coherence with the already superconducting
carriers,  the carriers in the boundary layer that is becoming superconducting are `carried along' by the
carriers in the layer right below, that are moving with speed $v_s$. However this argument has two problems: (1) it does not explain how the speed $v_s$ is generated in the first place, when the transition to 
superconductivity begins; (2) in the cylindrical geometry of Fig. (2b), the momentum (angular momentum in this case) of the current near the $s-n$ boundary increases
as the boundary moves out because its perimeter increases, so a simple momentum transfer from one layer to the next cannot account for this growth of total momentum.

Thus, we argue that this momentum (or angular momentum) of the normal carriers becoming superconducting has to originate  in an external source, not in the 
superconducting carriers themselves. The only conceivable external source is the magnetic field $H$. 
However, we  can conceive of no mechanism for $H$ to impart momentum in the $y$ direction to the carriers within
conventional BCS-London theory \cite{lorentz}.

\subsection*{(2) How is momentum conserved?}
There is no momentum of the electromagnetic field in the $y$ direction (since it is in direction $\vec{E}\times\vec{H}$) and assuming the process is sufficiently slow no momentum is
carried away by electromagnetic waves, so {\it the only way} to compensate for the change in momentum Eq. (40b) is to transfer it to the lattice.
If we assume for definiteness that the carriers have charge $q>0$ so they move in the same direction as the current, the lattice as a whole has to acquire momentum in the positive $y$
direction (to the left in Fig. 8) when the boundary layer in Fig. 8 becomes superconducting.

It may be argued that there is a  mechanism based on the electron-phonon interaction to account for such a momentum transfer to the lattice. This has not been  discussed in the literature, 
and we can conceive of no way
in which  the argument could be made. Furthermore there are a large number of materials classes for which it is generally believed that the electron-phonon interaction is not involved in the 
mechanism of superconductivity \cite{specialissue}, where superconductivity is believed to be caused by electron-electron interactions. How can carriers that bind into Cooper pairs through a non-phonon
mechanism interact with the lattice in such a way that they transfer the right amount of momentum, given by Eq. (40b), whose magnitude is determined by the magnetic field,  to the lattice?
We can conceive of no answer to this question within the conventional understanding  of superconductivity \cite{momentum}.

\subsection*{(3) How is the condensation energy converted into kinetic energy of the supercurrent?}
The energy of the carriers is lowered by condensation into the superconducting state. This energy is converted into the kinetic energy of the
carriers in the boundary layer becoming superconducting as given by Eq. (41). However the energy lowering, presumably associated with Cooper pairing, has no directionality. How does it get converted
into a kinetic energy that is associated with motion in one particular direction, i.e. the $(-y)$ direction in Fig. 8? Again we argue that the conventional BCS-London theory offers
no answer to this question.
 
In the absence of answers to these three questions we argue  that the Meissner effect remains unexplained within the conventional theory of superconductivity.

\section{What  is wrong with the conventional understanding}
The questions that we raise here have not been raised in the context of the conventional theory of superconductivity. In this section
we try to understand why.

Let us start by recalling the argument frequently used to make London's equation Eq. (14b)  plausible \cite{schrieffer,am}. Starting from Newton's equation for dissipationless motion of a charged fluid under an electric field $\vec{E}$ and using Faraday's law,
\bmath
\beq
\frac{\partial \vec{J}}{\partial t}=\frac{n_s q^2}{m}\vec{E}
\eeq
\beq
\frac{\partial}{\partial t}  (\vec{\nabla}\times\vec{J})=-\frac{n_s q^2}{m c}\frac{\partial \vec{H}}{\partial t}  ,
\eeq
\emath
and integrating Eq. (42b) in time, Eq. (14b) results ``outside of a constant of integration'' \cite{schrieffer}, with $\lambda_L$ given by
Eq. (21). However, integrating Eq. (42b) in time yields
\beq
\vec{\nabla}\times\vec{J}(t)-\vec{\nabla}\times\vec{J}(t=0)=-\frac{n_s q^2}{m c}(\vec{H}(t)-\vec{H}(t=0))
\eeq
where the $\vec{r}$ dependence is implicit. 
If a normal metal is cooled into the superconducting state in the presence of a spatially uniform magnetic field $\vec{H}_0$ through its interior,
the initial superfluid current  $\vec{J}(t=0)=0$ and Eq. (43) yields
\beq
\vec{\nabla}\times\vec{J}(\vec{r},t)=-\frac{n_s q^2 }{m c}(\vec{H}(\vec{r},t)-\vec{H}_0)
\eeq
which is $not$ London's equation Eq. (14b) . Quite the contrary, Eq. (44) implies that $\vec{J}(\vec{r},t)=0$ and $\vec{H}(\vec{r},t)=\vec{H}_0$ for all times
$t>0$, so that the magnetic field   remains unchanged inside the superconductor.

It is also argued that London's equation and the Meissner effect follow from BCS theory \cite{tinkham}. However this is only true to the extent that one $assumes$ that in
cooling a metal into the superconducting state the system will reach the state of minimum energy described by BCS theory, which
implies the existence of a macroscopic wavefunction 
\beq \Psi(\vec{r})=|\Psi(\vec{r})|e^{i\theta(\vec{r})}
\eeq
with a unique well-defined macroscopic phase $\theta(\vec{r})$ describing all the electrons in
 the superfluid. From Eq. (45)   it follows, with $n_s=|\Psi(\vec{r})|^2$ and assuming uniform $|\Psi(\vec{r})|$,  that the supercurrent is given by 
\beq
\vec{J}=\frac{n_sq}{m}(\hbar\vec{\nabla} \theta-\frac{q}{c}\vec{A})
\eeq
hence Eq. (14b) results from taking the curl. However this is a circular argument. Assuming that Eq. (45) is valid implies that
phase coherence has been established throughout the system, which $implies$ that the magnetic field has been expelled,
since no phase coherence can be established in the presence of a magnetic field. The question of $how$ phase coherence
is established, i.e. what are the dynamical processes that lead from the normal state with no phase coherence to the superconducting
state described by Eq. (45) and the magnetic field expelled, needs to be addressed to make this argument valid. Otherwise, in the absence of experimental
evidence, it could also be concluded that the BCS state described by Eq. (45) will never be attained by a system cooled below its
critical temperature in the presence of a magnetic field.

It is  argued that time-dependent Ginzburg Landau theory (TDGL) \cite{schmid,abrahams} describes the time evolution of the superconducting
order parameter $\Psi$ and in particular explains  the Meissner effect, as assumed in the treatments of refs. \cite{dorsey,goldenfeld}.
However, in TDGL it is $assumed$ that the time evolution of the order parameter is such that it will evolve towards its
equilibrium value that minimizes the Ginzburg-Landau (GL) free energy. In particular, that a generalized force exists, given by the functional derivative of the GL free energy
with respect to the order parameter, that drives the time evolution  of the order parameter, making it relax to its equilibrium value in roughly  the time
it takes a carrier to diffuse over a distance of a coherence length \cite{tinkham2}. 
This is an assumption that cannot be rigurously derived from microscopic BCS theory, unlike the equilibrium GL free energy that 
can under certain approximations. In the `derivations' of TDGL from BCS it is $assumed$ that the system will reach the BCS 
ground state. For example, in Ref. \cite{schmid} it is postulated that making the electron-electron attraction $g$
 time-dependent, i.e. $g\rightarrow e^{\delta t}g$, $\delta>0$, so that the interaction is zero for $t\rightarrow -\infty$ and attains its
 value $g$ at $t=0$, results in an order parameter $\Delta\rightarrow e^{\delta t}\Delta$. Again this assumes the answer, i.e. that
 the BCS state with expelled magnetic field will be attained starting from a normal metal in the presence of a magnetic field when the metal becomes
 superconducting, without proving that this is the case and without specifying the physical
 processes by which this occurs. 
 
 More specifically, within TDGL the current in the superconducting region is always given by Eq. (46), with $n_s=|\Psi(\vec{r},t)|^2$.  As a boundary
 layer becomes superconducting the current grows not because the speed of the carriers changes continuously from $v_n$ to $v_s$ but because the number of
 superconducting carriers $n_s$ increases continuously as  $|\Psi(\vec{r},t)|$ increases. The speed of the superconducting carriers is fixed, at the
 value $v_s=-(q/mc)A$ determined by the London equation. Thus, in this description the velocity of an individual carrier changes 
 $discontinuously$ in sign and magnitude from $v_n$ to $v_s$ when the carrier becomes part of the superconducting condensate
 described by $\Psi$. The question of momentum conservation is not even addressed. We argue that this is not a satisfactory description of the physics.

\section{What is required to explain the Meissner effect}

We have argued in the introduction   that the Meissner effect necessarily involves
motion of the normal-superconducting phase boundary, namely that the scenario of Fig. 1(a) that does not
involve motion of the phase boundary can be excluded on purely theoretical grounds. 
In order to answer  the questions posed in Sect. V we propose that it is necessary to assume
that {\it the motion of the phase boundary is associated with motion of charge}. In the conventional understanding of superconductivity
no motion of charge is associated with the motion of the phase boundary \cite{goldenfeld,dorsey}.

    \begin{figure}
 \resizebox{8.5cm}{!}{\includegraphics[width=6cm]{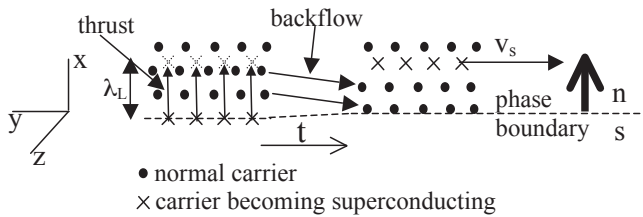}}
 \caption { As carriers become superconducting (crosses), they thrust forward in the $x$ direction a distance $\lambda_L$ and in the process acquire a speed $v_s$ 
 in the negative $y$ direction  (assuming positive carriers) due to the
 action of the Lorentz force. The normal carriers in the x-interval $(x_0, x_0+\lambda_L)$ undergo backflow in the negative x direction
 due to electrostatics. In the process they 
 acquire impulse in the positive $y$ direction due to the Lorentz force and transmit this impulse  to the lattice through normal scattering processes. 
  }
 \label{figure1}
 \end{figure} 

Assume  that as  carriers condense into the
superconducting phase they experience a sudden thrust in direction normal to the phase boundary and towards  the normal region, as shown
schematically in Fig. 9.  For simplicity
we assume $q>0$ for the moment. The Lorentz force acting on these carriers moving with 
velocity $\vec{v}_x$ in the positive $x$ direction 
\beq
\vec{F}_L=\frac{q}{c} \vec{v}_x\times\vec{H}=-\frac{q}{c}v_xH_c \hat{y}
\eeq
points in the negative $y$ direction, and if it is larger than the Faraday force $qE_y$ it will
impart motion to the carriers in the negative $y$ direction which is the direction of the current $J_y$ in the superconducting region.
The total force including the force from the Faraday field $E_y$ at the phase boundary (Eq. (33))  is
\beq
\vec{F}_L=q(E_y \hat{y}+ \frac{\vec{v}_x} {c} \times\vec{H})=-\frac{q}{c}(v_x-\frac{\partial x_0}{\partial t})H_c \hat{y}
\eeq
so we can ignore the Faraday field if $v_x>>\partial x_0/\partial t$. The speed in the y direction that these carriers acquire if they thrust forward a distance $\Delta x$ in a time
interval  $\Delta t$ is
\beq
v_y=\int_0^{\Delta t}   \frac{F_L}{m} dt=-\frac{q H_c}{c} \int_0^{\Delta t} v_x dt=-\frac{q H_c}{c} \Delta x
\eeq
and for $\Delta x=\lambda_L$
\beq
v_y= -\frac{q}{mc}\lambda_L H_c= -\frac{c}{4\pi \lambda_L}\frac{H_c}{qn_s}=v_s  .
\eeq
Therefore, the carriers becoming superconducting have  to thrust forward a distance $\lambda_L$ into the normal region to acquire the required speed Eq. (39b).

As these superconducting carriers are thrusting into the normal region in  the positive $x$ direction they will cause a counterflow of
charge from the normal region in the negative $x$ direction  due
to electrostatic forces, in order to restore charge uniformity, as shown in Fig. 9. The counterflowing carriers are still in the normal state, and experience scattering with
lattice imperfections and phonons. Through the Lorentz force Eq. (47) they acquire the same impulse in the 
positive $y$ direction (since $v_x<0$ for these carriers) as the thrusting superconducting carriers acquired in the negative $y$ direction, and they transmit this impulse
  to the lattice as a whole through normal scattering processes. This accounts for the momentum conservation question discussed
in   section V. 

Finally, the third question raised in  section V about the conversion of condensation energy into kinetic energy of
motion  in a definite
direction   is also answered by this process.    The energy for the
thrusting forward is provided by the condensation energy, the direction of the current carrying the kinetic energy
is determined by the   Lorentz force Eq. (47) acting on the carriers thrusted in the $+x$ direction.

      \begin{figure}
 \resizebox{8.5cm}{!}{\includegraphics[width=6cm]{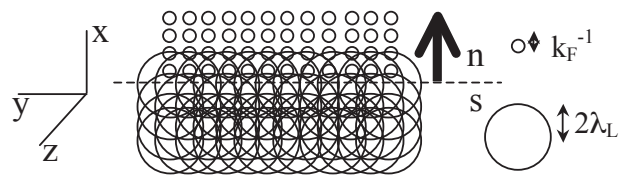}}
 \caption {Schematic depiction of the orbits of the carriers in the theory of hole superconductivity. Normal state carriers have small non-overlapping orbits  of radius $k_F^{-1}$, 
 superconducting state carriers have large overlapping orbits of radius $2\lambda_L$. As normal carriers become superconducting their orbits expand,  they acquire azimuthal velocity $v_\phi$ due to the magnetic field, and charge is transferred from the 
 superconducing to the normal side of the phase boundary.}
 \label{figure1}
 \end{figure}

\section{Explanation of the Meissner effect within the theory of hole superconductivity}

The theory of hole superconductivity   \cite{holesc} proposes a concrete realization of the physical processes described in the previous section. 
 Within this theory, charge carriers change their character from hole-like to electron-like when they pair and condense into the 
 superconducting state \cite{holeelec}. In a semiclassical description, electronic orbits expand from a   microscopic radius $k_F^{-1}$ ($k_F=$Fermi wavevector) to mesoscopic radius $2\lambda_L$ \cite{orbitexp}. 
 This orbit expansion is driven by lowering of kinetic energy \cite{kinenergy}, and has associated with it
 expulsion of negative charge from the interior   to the surface of superconducting bodies \cite{chargeexp}. 
 In the geometry considered here, it leads to transfer of negative charge from the superconducting into the normal region at the phase boundary, 
 as shown schematically in Fig. 10.

 That the system in the normal and superconducting states can be characterized by orbits of radius $k_F^{-1}$ and $2\lambda_L$ respectively can be seen from the magnetic susceptibility.
 The Larmor diamagnetic susceptibility for electrons of density $n$ per unit volume in orbits of radius $r$ perpendicular to the magnetic field is
 \beq
\chi_{Larmor}(r)=-\frac{n e^2}{4m_e c^2} r^2
\eeq
This yields the Landau diamagnetic susceptibility of the normal state when the orbits have radius $k_F^{-1}$:
\beq
\chi_{Larmor}(r=k_F^{-1})=-\frac{1}{3}\mu_B^2g(\epsilon_F)
\eeq
with $g(\epsilon_F)=3n/2\epsilon_F$ the density of states and
$\mu_B=|e|\hbar/2m_ec$, 
and perfect diamagnetism when the orbits have radius $2\lambda_L$
\beq
\chi_{Larmor}(r=2\lambda_L)=-\frac{n e^2}{4m_e c^2} (2\lambda_L)^2=-\frac{1}{4\pi} .
\eeq
The fact that electrons occupying  large orbits can account for the perfect diamagnetism of superconductors was pointed
out by Slater \cite{slater} and others long ago.

As  the orbit expands to radius $2\lambda_L$  in the presence of a magnetic field, the carrier in the orbit acquires an azimuthal velocity \cite{azimuth}
  \beq
 v_\theta=-\frac{q \lambda_L}{mc}H_c
 \eeq
 due to the action of the Lorentz force. This is precisely the speed Eq. (39b) (using Eq. 21), and thus provides a
 $dynamical$ explanation of the Meissner effect \cite{orbitexp}.   Because the electrons have negative charge the azimuthal motion acquired is in the
 counterclockwise direction. The superposition of these orbits gives rise to net electronic charge motion in the $+y$ direction, corresponding to current $J_y$ in the
 negative $y$ direction as given by Eq. (31). The rate of radial expansion is given by $v_x$ defined in the previous section,
 which can be plausibly assumed to be much larger than $\partial x_0/\partial t$. As the orbits at the phase boundary
 expand, they will enclose normal small orbits right above them (Fig. 10). These normal electrons will be pushed in the negative $x$ direction due to electrostatic interactions,
 since the expanding orbits carry negative charge with them that extends into the region of the normal phase right above the phase boundary. This accounts for the backflow 
 discussed in the previous section.  
 
 Thus, the theory of hole superconductivity provides a realization of the conditions discussed in the previous section that are needed to account for 
 energy and momentum conservation as the phase boundary advances into the normal region   in the presence of a magnetic field.
 
\section{Cylindrical geometry}

        \begin{figure}
 \resizebox{6.5cm}{!}{\includegraphics[width=6cm]{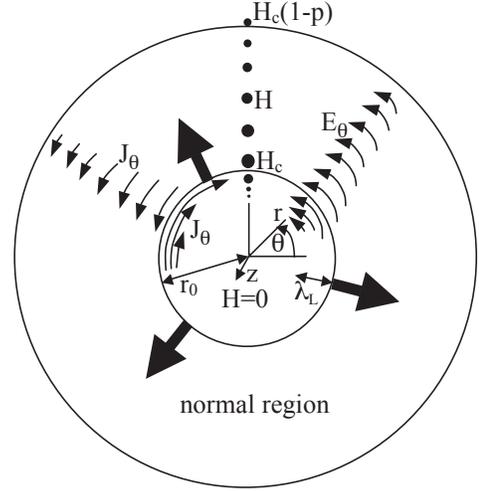}}
 \caption {Superconducting region expanding from the center in a cylindrical geometry. $0<r<r_0$ is the   superconducting region,
 $r_0<r<R$ is the normal region. The magnetic field points out of the paper in the $\hat{z}$ direction. }
 \label{figure1}
 \end{figure} 

Next we consider the growth of the superconducting phase in a cylindrical geometry as shown in Fig. 11. Here, the paradox
associated with momentum conservation (angular momentum in this case) is even more apparent than in the planar case
discussed earlier.

The magnetic field points along the axis of the cylinder ($\hat{z}$ direction), and the boundary between the superconducting and normal phase is given by radius $r=r_0(t)$, 
so that the system is superconducting for $0<r<r_0$ and normal for $r_0<r<R$, with $r_0(t=0)=0$ and
$r_0(t_0)=R$, with $t_0$ the time it takes for the magnetic field to be expelled, to be determined.  The induced Faraday field as the superconducting phase expands from
the center, $E_\theta(r)$, points in the $+\hat{\theta}$ direction, generating an azumuthal current $J_\theta$
{\it in the normal region}  that creates a magnetic field
in the $+\hat{z}$ direction that tries to restore the magnetic field being expelled. The analogous of Eqs. (2) for the cylindrical geometry are
\bmath
\beq
\vec{\nabla}\times \vec{H}=\frac{4\pi}{c}\vec{J}==>
\frac{\partial H}{\partial r}=-\frac{4\pi}{c}J_\theta
\eeq
\beq
\vec{\nabla}\times \vec{E}=-\frac{1}{c}\frac{\partial\vec{H}}{\partial t}==>
\frac{1}{r}\frac{\partial (r E_\theta)}{\partial r}=-\frac{1}{c}
\frac{\partial H}{\partial t}
\eeq
\beq
\vec{J}=\sigma \vec{E}==> J_\theta=\sigma E_\theta
\eeq
\emath
and the boundary condition analogous to Eq. (3a) is
\beq
E_\theta(r_0)=\frac{H_c}{c} \frac{\partial r_0}{\partial t}  .
\eeq
Replacing $E_\theta$ by Eq. (55c) and using Eq. (55a) the condition Eq. (56) is
\beq
\frac{\partial H}{\partial r})_{r=r_0}=-\frac{4\pi\sigma}{c^2}\frac{\partial r_0}{\partial t}
\eeq
and combining Eqs. (55) yields
\beq
\frac{1}{r} \frac{\partial }{\partial r}(r\frac{\partial H}{\partial r})=\frac{4\pi\sigma}{c^2}\frac{\partial H}{\partial t}
\eeq
similar to Eq. (4) for the planar case. The boundary conditions are
\bmath
\beq 
H(r=R)=H_c(1-p)
\eeq
\beq
H(r=r_0)=H_c
\eeq
\emath
and Eq. (56).

In the planar case, separation of variables was achieved by defining the variable $y=x/x_0$, and the boundary conditions were
given at $y=0$ ($x=0$) and $y=1$ ($x=x_0$). This doesn't work in the present case. If we define $y=r/r_0$ the boundary conditions
are given at $y=1$ and $y=R/r_0$, which are not time-independent. Time-independent boundary conditions can be achieved
by defining $r=(r_0-R)y+R$, so that the conditions Eq. (59a) and (59b) are given at $y=0$ and $y=1$ respectively, but in terms of
this variable Eq. (58) does not separate as Eq. (7) did.

Eq. (58), with boundary conditions Eqs. (57), (59) can of course be solved numerically. Instead, we discuss here an approximate
analytic solution, following Pippard\cite{pip}, which becomes exact in the limit $p\rightarrow 0$. To zeroth order in $p$  we may assume 
that the magnetic field in the
normal region is given by $H=H_c$, which  yields for the induced electric field in the normal region
\beq
E_\theta (r)=\frac{H_c}{c}\frac{r_0}{r}\frac{\partial r_0}{\partial t} .
\eeq
Using Eqs. (55c) and Eq. (55a) integrated from $r_0$ to $R$ yields
\beq
r_0\frac{\partial r_0}{\partial t} ln\frac{R}{r_0}=\frac{pc^2}{4\pi\sigma}
\eeq
and performing the time integration from $0$ to $t$ with $r_0(t=0)=0$ and $r_0(t)=r_0$ yields
\bmath\beq
(\frac{r_0}{R})^2[1+2 ln\frac{R}{r_0}]=\frac{t}{t_0}
\eeq
with
\beq
t_0=\frac{\pi\sigma R^2}{p c^2}
\eeq
\emath
the total time for the magnetic field to be expelled from the cylinder.

        \begin{figure}
 \resizebox{8.5cm}{!}{\includegraphics[width=6cm]{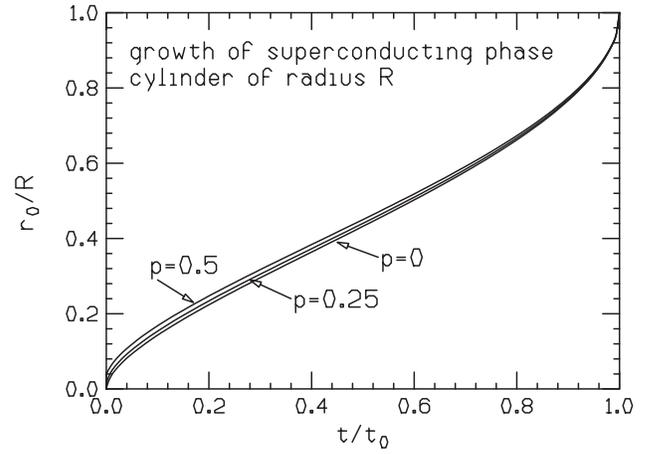}}
 \caption {Radius of the superconducting region $r_0(t)$ versus time t }
 \label{figure1}
 \end{figure}

 These results are exact in the limit $p\rightarrow 0$. In the next order of approximation we assume a linear interpolation for the magnetic field 
 between $r=r_0$ and $r=R$. This approximation, for the planar case, yields the dashed lines shown in Fig. 4, corresponding to
 $\alpha=3/(3+p)$ for the `n growing' case and $\alpha=3/(3-p)$ for the `s growing' case. It can be seen in Fig. 4 that that they closely follow
 the exact solutions for the planar case, hence we may similarly assume that the results will be close to the exact results in the
 cylindrical case. Hence we assume
\beq
H(r)=H_c(1-p\frac{r-r_0}{R-r_0})
\eeq
varying linearly from $H_c$ at $r=r_0$ to $H_c(1-p)$ at $r=R$. This 
 leads instead of Eq. (60) to
\beqn
E_\theta(r)&=&\frac{H_c}{c}\frac{r_0}{r}\frac{\partial r_0}{\partial t} [1+\frac{p}{r_0(R-r_0)^2}\times \nonumber \\
& &[\frac{r^3}{3}-\frac{r_0^3}{3}+\frac{R}{2}(r_0^2-r^2)]]
\eeqn
Using Eqs. (55a) and (55c) we find that the time required for field expulsion now is slightly less than given by Eq. (62b),
\beq
t_0=\frac{\pi\sigma R^2}{p c^2}(1-\frac{4}{9}p)  .
\eeq
The equation giving the time evolution of $r_0$ in this approximation is lengthy and will not be reproduced here. In Fig. 12 we
plot the time evolution of $r_0$ for various values of $p$ obtained within this approximation, the curve labeled $p=0$ corresponds to Eq. (62a). It can be seen
that the results obtained for the time evolution for different p's are very similar provided the time is scaled according to the relation Eq. (65).
From this we conclude that the exact results obtained from numerical integration of Eq. (58) would be very similar to the results predicted by the
lowest order approximation Eq. (62).

These results are valid in the normal phase. In the lowest order approximation, the current at the n-s phase boundary on the normal
side is given  (from Eqs. (60) and  (55c) by 
\beq
J_\theta(r_0)=\frac{cp}{4\pi r_0 ln\frac{R}{r_0}}H_c
\eeq
and flows in the positive $\theta$ direction (counterclockwise).

In the superconducting phase the fields and current are governed by Ampere's law and  London's equation \cite{londonbook} 
\bmath
\beq
\vec{\nabla}\times \vec{H}=\frac{4\pi}{c}\vec{J}==>
\frac{\partial H}{\partial r}=-\frac{4\pi}{c}J_\theta
\eeq

\beq
\vec{\nabla}\times \vec{J}=-\frac{c}{4\pi\lambda_L^2}\vec{H}==>
\frac{1}{r}\frac{\partial (r J_\theta)}{\partial r}=-\frac{c}{4\pi \lambda_L^2}H  .
\eeq
\emath 
Combining Eqs. (67a) and (67b) yields 
\beq
\frac{1}{r}\frac{\partial}{\partial r}(r\frac{\partial H}{\partial r})=\frac{1}{\lambda_L^2}
\eeq
and the solution with boundary condition $H(r_0)=H_c$ is \cite{laue,bardeen56}
\beq
H=H_c\frac{I_0(r/\lambda_L)}{I_0(r_0/\lambda_L)}
\eeq
with $I_0(x)$ a Bessel function of imaginary argument, with asymptotic behavior 
\beq
I_0(x)\rightarrow \frac{e^x}{\sqrt{2\pi x}}
\eeq
for large $x$. Hence, assuming $r_0>>\lambda_L$ the magnetic field in the region $r<r_0$ is given by
\beq
H(r)=H_ce^{(r-r_0)/\lambda_L}\sqrt{\frac{r_0}{r}}
\eeq
and from Eq. (67a) the current in the superconducting region is given by
\beq
J_\theta(r)=-\frac{c}{4\pi\lambda_L}H_ce^{(r-r_0)/\lambda_L}\sqrt{\frac{r_0}{r}}(1-\frac{\lambda_L}{2r})
\eeq
and flows in the negative $\theta$ direction (clockwise). Hence at the n-s phase boundary on the superconducting side the current is (for $r_0>>\lambda_L$)
\beq
J_\theta=-\frac{c}{4\pi\lambda_L}H_c
\eeq
i.e. it is much larger in magnitude than the current in the normal side (Eq. (66)) and flows in opposite direction. Thus, just like in the case of the planar interface there is a large 
discontinuity in magnitude  and a change in sign of  the current  at the phase boundary, as shown schematically in Fig. 11.

The electric field in the superconducting region is obtained from Faraday's law Eq. (55b) and  the expression for the magnetic field Eq. (71)  
 \beq
 \frac{1}{r}\frac{\partial}{\partial r}(rE_\theta(r))=\frac{H(r)}{c\lambda_L}\frac{\partial r_0}{\partial t}(1-\frac{\lambda_L}{2r_0})
 \eeq
 with  the variation of $r_0(t)$ with
$t$ given by Eq. (61). Assuming the dominant variation comes from the exponential behavior in Eq. (71), which is 
the case  for $r>>\lambda_L$, yields for the electric field in the 
superconducting region
\beq
E_\theta(r)=-\frac{H_ce^{(r-r_0)/\lambda_L}}{c \lambda_L}\sqrt{\frac{r_0}{r}} \frac{\partial r_0}{\partial t} (1-\frac{\lambda_L}{2r}_0)
\eeq
so that for $\lambda_L<<r_0, r$
\beq
\frac{\partial J_\theta(r)}{\partial t}=\frac{c^2}{4\pi \lambda_L^2}E_\theta(r)
\eeq
holds as in the case of the planar interface, describing the decelaration of the carriers in the superconducting region due to the electric field that points in direction opposite
to the superconducting current flow.

\subsection*{Angular momentum}

As the superconducting region expands, the current carriers in the superconducting region acquire increasing angular momentum, both because the angular momentum of each carrier of the supercurrent near $r_0$ increases proportionally to $r_0$ 
and because the number of carriers within $\lambda_L$ of the phase boundary grows as the perimeter of the circle of radius $r_0$
increases. It is found 
experimentally\cite{kik,pry,doll} that the mechanical angular momentum carried by the Meissner current for a cylinder of radius
$R$ and height $h$ with applied magnetic field $H$ parallel to the cylinder axis has magnitude
\beq
L_e=\frac{m_ec}{2|e|} R^2hH
\eeq
and points in direction $parallel$ to the applied magnetic field. In Eq (77) $m_e$ is the $bare$ electron mass. This relation
holds for all superconductors \cite{kik,pry,doll} and results if the
carriers in the Meissner current are bare electrons of mass $m_e$ and negative charge $e$ \cite{lenz}.

We can ignore the angular momentum carried by the eddy currents generated in the normal phase,
because it is much smaller than the angular momentum carried by the superconducting carriers. When the phase boundary is at radius $r_0$
the total electronic angular momentum is then
\beq
\vec{L}_e(t)=-\frac{m_ec}{2e}hH_cr_0(t)^2 \hat{z}
\eeq
pointing in the positive $\hat{z}$ direction. By conservation of angular momentum we have to assume that the body as a whole acquires
an equal and opposite angular momentum
\beq
\vec{L}_{body}(t)= \frac{m_ec}{2e}hH_cr_0(t)^2 \hat{z}
\eeq
and given the time evolution of $r_0$, Eq. (62a), we have
\beq
\vec{L}_{body}(t)=\frac{m_ec}{2\pi e} H_c V \frac{t}{t_0}\frac{1}{1+2 ln\frac{R}{r_0}} \hat{z}
\eeq 
with $V$ the volume of the body. This corresponds to an angular velocity of the body
\beq
\omega(t)=\frac{m_ec}{\pi |e| \rho} \frac{H_c}{R^2}  \frac{t}{t_0}\frac{1}{1+2 ln\frac{R}{r_0}}  
\eeq
with $\rho$ the body's mass density. Thus, as the superconducting phase grows, a macroscopic angular momentum and angular 
velocity of the body as a whole is generated that grows approximately linearly with time, given by Eqs. (80) and (81).
 The torque acting on the body that causes its rotation is, from Eq. (79)
\beq
\vec{\tau}=\frac{d \vec{L}_{body}}{dt}= \frac{m_ec}{e}hH_c r_0 \frac{d r_0}{dt} \hat{z}
\eeq
which we can write using Eqs. (61) and (62b) as
\beq
\vec{\tau}= \frac{m_ec}{e} \frac{H_c}{4t_0} \frac{h R^2}{ln \frac{R}{r_0}} \hat{z}
\eeq
For example, for a sample of $R=1cm$, $h=5cm$, $H_c=200G$, if it takes $t_0=1s$ to expel the magnetic field, $\tau=1.4\times 10^{-5} gr cm^2/s^2$
when $r_0=R/e=0.37R$.

The conventional theory of superconductivity provides no explanation for the origin of this macroscopic torque, nor for how the  electronic angular momentum and ionic
counter-angular momentum are generated and grow as the superconducting phase grows. In contrast, the theory of hole superconductivity
provides an explanation according to the physics discussed in Sect. VIII, shown schematically in Fig. 13.
In a cylinder of radius $r_0$ and height $h$, the total angular momentum due to carriers of density $n_s$ moving in orbits of
radius $2\lambda_L$ with orbital speed $v_\theta$ is
\bmath
\beq
L_e=[m_ev_\theta (2\lambda_L)]n_s (\pi r_0^2)h
\eeq
which can be rewritten as
\beq
L_e=[m_ev_\theta r_0]n_s(2\pi r_0\lambda_L)h .
\eeq
\emath
In the first form, it describes the aggregate angular momenta of carriers of density $n_s$ occupying an area $\pi r_0^2$, each in an orbit
of radius $2\lambda_L$ with angular momentum $m_e v_\theta (2\lambda_L)$. In the second form, it describes the
angular momentum from carriers in a ring of radius $r_0$ and thickness $\lambda_L$, i.e. the carriers of the Meissner current when the
superconducting region has radius $r_0$. Clearly these two descriptions are equivalent, since superposing the $2\lambda_L$ orbits the
internal velocities cancel out and only the surface current remains. The speed of the carriers at radius $r_0$ is given by Eq. (73).
Assuming the carriers are bare electrons of density $n_s$, as implied by the experiments \cite{kik,pry,doll}, Eq. (73) with
$J_\theta=en_s v_\theta$ yields
\beq
v_\theta=-\frac{c}{4\pi \lambda_L n_s e}H_c
\eeq
and from Eq. (82)
\beq
L_e=-\frac{m_e c}{2e}r_0^2 h H_c
\eeq
in agreeement with Eq. (78).

       \begin{figure}
 \resizebox{8.5cm}{!}{\includegraphics[width=6cm]{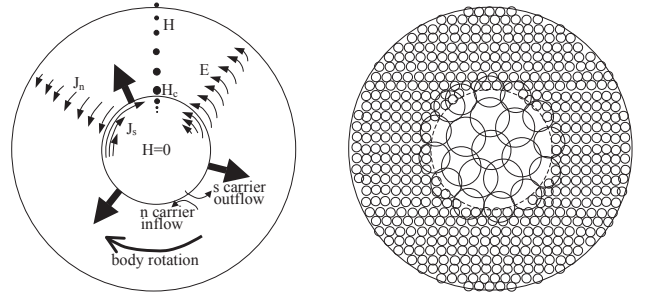}}
 \caption {Left panel: pattern of currents and fields in  cylindrical geometry. The magnetic field points out of the paper. As the superconducting phase grows, the boundary supercurrent acquires increasing angular momentum. Assuming
 the current is carried by negative charge carriers (electrons),
 as shown by experiment \cite{kik,pry,doll},
  the body as a whole  rotates in the clockwise direction to conserve angular momentum. The right panel depicts the expanded orbits in the superconducting
 region that explain the process within the theory of hole superconductivity.}
 \label{figure1}
 \end{figure}

The generation of angular momentum occurs as follows:  as the
phase boundary moves out, the orbits of normal carriers becoming superconducting carriers  expand from   microscopic radius $k_F^{-1}$ to
radius $2\lambda_L$. In the presence of magnetic field $H_c$, the electron in the expanding orbit
experiences a Lorentz force and acquires velocity $v_\theta$ in the positive $\hat{\theta}$ direction given by
Eq. (85), or equivalently Eq (54) \cite{azimuth}, and hence angular momentum in the $+\hat{z}$ direction. The sum of these dynamically generated
angular momenta yields the total electronic angular momentum Eq. (82). Furthermore, as the orbits at the phase boundary
expand, there is an outflow of negative charge from the superconducting into the  normal region that gives rise to a radial backflow of normal carriers as discussed in Sect. VII.
The carriers flowing inward acquire angular momentum in the $-\hat{z}$ direction through the action of the Lorentz force and
transmit this angular momentum to the lattice by collisions. In this way, as the phase boundary moves out, electrons acquire increasing
angular momentum in the $+\hat{z}$ direction and the body acquires the compensating angular momentum in the
$-\hat{z}$ direction given by Eqs. (78) and (79).

We argue that there is no way to explain the dynamics of this process of angular momentum generation in the absence of the
radial outflow and inflow of charge discussed here.

\section{Summary and discussion}
In this paper we have calculated the electromagnetic fields and currents in the process where the normal phase grows into the superconducting phase and
in the reverse process where the superconducting phase grows into the normal phase, in the
presence of a magnetic field, assuming for simplicity a planar interface, and also for a cylindrical geometry. 
We have also obtained the speed of motion of the phase boundary in these 
processes.

The results obtained are easily understood for the  process where the normal phase grows. However, for the reverse process, which correspond
to the Meissner effect, we have argued that the results cannot be understood within the conventional understanding of superconductivity, i.e.  the 
London equation, BCS-Eliashberg theory and time-dependent Ginzburg-Landau theory. This conventional framework describes pairs of carriers that change
their center of mass velocity discontinuously in both sign and magnitude as they bind into a Cooper pair and condense into the superconducting state in the presence of a magnetic field, without 
providing a physical explanation for how this change of velocity, in direction opposite to that dictated by the force exerted by the Faraday electric field, occurs, 
 and without explaining
 how the change of momentum is compensated so that the physical law of momentum
conservation is not violated. It is expected that the change in momentum is compensated by momentum transferred to the lattice as a whole, but no mechanism for such
transfer is provided and it is not clear how this would occur, particularly for superconductivity mechanisms that do not involve the electron-phonon interaction. 

For a cylindrical geometry, the question of momentum conservation is even more puzzling than in the planar geometry, because 
the  momentum grows with time rather than staying constant in total magnitude as in the planar geometry. We calculated the rate of growth of
angular momentum of the conducting carriers and of the body as a whole, and the magnitude of the angular velocity of the body and of the torque
acting on the body as the magnetic field is expelled. These are macroscopic observables for which we argue  there is no explanation within the
conventional theory of superconductivity.

It is true that for a microscopic quantum process, such as the transition of an electron from a $2p$ to a $1s$ atomic orbital with emission of a photon, we do not object to the fact that the electron
changes its velocity discontinuously. However, the superconductor is a macroscopic system, albeit quantum, and in the spirit of Bohr's correspondence principle one would expect
that its behavior should be understandable without invoking processes that seem to fly in the face of macroscopic physical laws, as the TDGL description does. And even for 
microscopic atomic processes there is an account of energy and momentum (or angular momentum) conservation (angular momentum is carried away by the photon in the
above example), while no explanation of how momentum or angular momentum conservation
is satisfied in the Meissner effect is provided by the TDGL formalism. Finally, the fact that one state of a system has a lower energy than another state does not predict that the system will evolve from the high energy to the low energy state, even in the microscopic realm. For example, two $Be$ atoms will not spontaneously fuse to form an $O$ atom.

Instead, we have  argued that the Meissner effect $can$ be understood by   making assumptions
that are not part of the conventional understanding of superconductivity, namely that there is {\it motion of 
charge in direction normal to the phase boundary} associated with the motion of the phase boundary.
More specifically, that there is motion of ``superconducting charge'', that undergoes no scattering, 
from the superconducting into the normal region, and backflow of ``normal charge'', that does undergo
scattering, in the opposite direction, as the phase boundary advances into the normal region.
To understand the Meissner effect it is not necessary to assume a specific sign for the charge involved in the
charge flow perpendicular to the phase boundary. However within the
theory of hole superconductivity the sign is uniquely defined: it is {\it negative charge} that moves from the
superconducting into the normal region \cite{ehasym}.

Note that the proposed processes resemble processes known to occur in superfluid $^4He$. 
The process   where a superconducting region expands into a normal region
will occur  
if the superconducting region  is supercooled  relative to the normal region. In superfluid $^4He$, it is well known that there is flow of superfluid from colder to
warmer regions and counterflow of normal fluid (fountain effect), precisely what is required to explain the Meissner effect as
discussed here.   We have proposed that this commonality of behaviours occurs because both for superfluid $^4He$ and for superconductors described by the theory of hole
superconductivity, the transition into the superfluid or superconducting state is driven by lowering of {\it kinetic} energy  \cite{supersuper}, rather than lowering of potential energy as predicted by the conventional
theory.

The Meissner effect has been known for 82 years, and it is generally believed that it is explained within
the conventional BCS-London theory of superconductivity. We have argued  here  that this is not the case, and that the Meissner effect can only be understood if there is   charge motion in the
direction of motion of the phase boundary, which is not predicted by the conventional theory but is predicted by  the alternative   theory of hole superconductivity.
Thus we argue that the Meissner effect remains unexplained within the conventional theory, and hence that
the conventional theory in its present form can only apply to superconductors that do not exhibit the Meissner effect,
in other words that  it does not describe any of the currently known superconducting materials \cite{specialissue}.

\end{document}